\documentclass[aps,prl,preprint,groupedaddress]{revtex4}

\usepackage{epsfig}

\begin{document}


\preprint{DESY 05-146}
\preprint{MZ-TH/05-16}
\preprint{hep-ph/0508129}
\title{Reconciling open charm production at the Fermilab Tevatron with QCD}
\author{B.A. Kniehl,${}^1$ G. Kramer,${}^1$ I. Schienbein,${}^1$ and
H. Spiesberger${}^2$}
\affiliation{${}^1$ {II}.~Inst.\ f\"ur Theor.\ Physik, Univ.\
Hamburg, Luruper Chaussee 149, 22761 Hamburg, Germany\\
${}^2$ Institut f\"ur Physik, Johannes-Gutenberg-Universit\"at,
Staudinger Weg 7, 55099 Mainz, Germany}
\date{\today}

\begin{abstract}
We study the inclusive hadrodroduction of $D^0$, $D^+$, $D^{*+}$, and $D_s^+$ 
mesons at next-to-leading order in the parton model of quantum chromodynamics
endowed with universal non-perturbative fragmentation functions (FFs) fitted
to $e^+e^-$ annihilation data from CERN LEP1.
Working in the general-mass variable-flavor-number scheme, we resum the
large logarithms through the evolution of the FFs and, at the same time,
retain the full dependence on the charm-quark mass without additional
theoretical assumptions.
In this way, the cross section distributions in transverse momentum recently
measured by the CDF Collaboration in run~II at the Fermilab Tevatron are
described within errors.
\end{abstract}
\maketitle


Recently, there has been much interest in the study of charmed-hadron ($X_c$)
production at hadron colliders, both experimentally and theoretically.
The CDF Collaboration measured the differential cross sections $d\sigma/dp_T$
for the inclusive production of $D^0$, $D^+$, $D^{*+}$, and $D_s^+$ mesons
(and their antiparticles) in $p\bar{p}$ collisions at the Fermilab Tevatron
(run II) as functions of transverse momentum ($p_T$) in the central rapidity
($y$) region \cite{CDF}.
Unfortunately, the most advanced theoretical predictions available so far
\cite{BK,CN}, based on quantum chromodynamics (QCD) at next-to-leading order
(NLO), consistently undershoot all the $D^0$, $D^+$, and $D^{*+}$ data by
significant amounts, as is evident from Fig.~3 of Ref.~\cite{CDF}, while no
predictions for $D_s^+$ mesons exist yet.
It is presently an open question if this discrepancy is related to an
experimental problem, a technically deficient QCD prediction, or the
appearance of new physics beyond the standard model.
Such a situation is familiar from inclusive bottom-flavored-hadron ($X_b$)
production in run~I, where a long-standing discrepancy between CDF data
\cite{ABE} and certain NLO predictions of QCD were, in fact, interpreted as an
indication for low-energy supersymmetry \cite{BER}.
It is, therefore, an urgent task to deepen our understanding of the inclusive
hadroproduction of charmed hadrons on the basis of QCD in order to render the
theoretical predictions as reliable as possible, so as to establish a sturdy
anchor for new-physics searches.
This is even more important in view of future physics at the CERN Large Hadron
Collider, where the continuum production of charmed hadrons will provide
important backgrounds for numerous new-physics signals.
This is the main motivation of this letter.
Moreover, we provide the first NLO prediction for the CDF $D_s^+$ data
\cite{CDF}.
Prior to explaining our improved theoretical framework and describing our
updated input, for the reader's quick orientation, we present a brief survey
of the various NLO approaches adopted so far in the literature.

In the so-called massless scheme, also known as zero-mass
variable-flavor-number (ZM-VFN) scheme \cite{CG,BKK}, the conventional parton
model approach implemented in the modified minimal subtraction
($\overline{\rm MS}$) scheme is adopted, assuming that the charm ($c$) quark
can be treated as massless, although its mass $m$ is certainly larger than the
asymptotic scale parameter $\Lambda_{\rm QCD}$. 
In this approach, the $c$ quark occurs as an incoming parton, leading to
contributions in addition to those where it is produced by an incoming gluon
$g$ or a light $u$, $d$, or $s$ quark.
The $c$ quark fragments into the charmed hadron similarly as the gluon and the
light quarks with a fragmentation function (FF), which must be known from
other processes.
The well-known factorization theorem provides a straightforward procedure for
systematic higher-order perturbative calculations. 
Due to the assumption that $m=0$, the predictions are reliable only for large
values of $p_T$, with $p_T \gg m$, where powers of $m^2/p_T^2$ can be
neglected. 
This approach has the advantage, however, that the potentially large
logarithms of the type $\ln(p_T^2/m^2)$ are absorbed into the $c$-quark parton
distribution functions (PDFs) of the colliding hadrons and into the FF for the
transition $c\to X_c$.
These logarithms are resummed through the
Dokshitzer-Gribov-Lipatov-Altarelli-Parisi (DGLAP) evolution equations. 
Still, $m$ appears in the initial conditions of the $c$-quark PDF and FF. 
In this respect, the $c$-quark PDF and FF differ from the PDFs of the gluon
and the light quarks.

In the so-called massive scheme, also called fixed-flavor-number (FFN) scheme 
\cite{BKNS}, the number of active quark flavors in the initial state is fixed
to $n_f=3$, and the $c$ quark only appears in the final state.
The physical value of $m$ is explicitly taken into account together with
the variable $p_T$, as if the two were of the same order. 
In this scheme, $m$ acts as a cutoff for the initial- and final-state
collinear singularities. 
However, in NLO, terms proportional to $\ln(p_T^2/m^2)$ arise at large $p_T$
values from collinear gluon emission off $c$ quarks or from almost collinear
branchings of gluons into $c\bar{c}$ pairs. 
For $p_T\gg m$, these terms spoil the convergence of the perturbative series. 
The FFN approach with $n_f=3$ is thus limited to a narrow $p_T$ range,
reaching up to a few times $m$. 
The advantage of this scheme is that the $m^2/p_T^2$ power terms are fully
taken into account.

Obviously, the ZM-VFN and FFN schemes are valid in complementary $p_T$
regions, and is desirable to combine them in a unified approach that enjoys
the virtues of both schemes and, at the same time, is bare of their flaws,
i.e.\ one that resums the large logarithms, retains the full finite-$m$
effects, and preserves the universality of the FFs.
This is vital for a reliable and meaningful interpretation of the CDF data
\cite{CDF}, which mostly populate the transition region.
A first attempt to implement such an interpolating scheme is the so-called
fixed-order next-to-leading-logarithm (FONNL) scheme, in which the conventional
cross section in the FFN approach is linearly combined with a suitably
modified cross section in the ZM-VFN approach with perturbative FFs,
using a suitable $p_T$-dependent weight function \cite{CN,CGN}. 
In both finite-mass approaches, FFN and FONNL, the FFN cross sections are
convoluted with a non-perturbative $c$-quark FF, adjusted to $e^+ e^-$ data,
that is not subject to DGLAP evolution.

Here, we wish to advocate an approach that is much closer in spirit to the
ZM-VFN scheme, but keeps all $m^2/p_T^2$ power terms in the hard-scattering
cross sections, namely the general-mass variable-flavor-number (GM-VFN)
scheme, which has recently been elaborated for the photo- \cite{KS} and
hadroproduction \cite{Kniehl:2004fy,Kniehl:2005mk} of heavy-flavored hadrons.  
In this approach, one starts from the $p_T\gg m$ region and absorbs the large
logarithms $\ln(p_T^2/m^2)$ into the $c$-quark PDF of the incoming hadrons
and the FF for the $c\to X_c$ transition.
After factorizing the $\ln m^2$ terms, the cross section is infrared safe in
the limit $m\to0$, and $n_f=4$ is taken in the strong-coupling constant
$\alpha_s$ and the DGLAP evolution equations.  
The remaining $m$ dependence, i.e.\ the $m^2/p_T^2$ power terms, is retained
in the hard-scattering cross sections.
These terms are important in the intermediate $p_T$ region, where $p_T\agt m$,
and are expected to improve the precision of the theoretical predictions.
The large logarithms are absorbed into the PDFs and FFs by subtraction of the
collinearly (mass) singular terms. 
However, in order to define a unique factorization prescription, one also has
to specify non-singular terms.
This is done by requiring that, in the limit $p_T\to\infty$, the known ZM-VFN
hard-scattering cross sections are recovered. 
To achieve this, subtraction terms are derived by comparing the FFN theory in
the limit $m\to0$ with the ZM-VFN theory, implemented in the
$\overline{\rm MS}$ factorization scheme.
This matching procedure is useful, since all commonly used $c$-quark PDFs and
FFs are defined in the ZM-VFN scheme. 
The latter can then be used consistently together with hard-scattering cross
sections calculated in the GM-VFN scheme. 
The derivation of the subtraction terms is described in
Ref.~\cite{Kniehl:2005mk}.

We note that our implementation of the GM-VFN scheme is similar to the
Aivazis-Collins-Olness-Tung (ACOT) \cite{acot} scheme formulated for the
initial state of fully inclusive deep-inelastic scattering.
The extension of this scheme to the inclusive production of heavy partons was
considered in Ref.~\cite{Olness:1997yc}, where the resummation of the
final-state collinear logarithms was only performed to leading order (LO) and
parton-to-hadron FFs were not included.
A comprehensive discussion of the differences between our approach and
Ref.~\cite{Olness:1997yc} concerning the collinear subtraction terms may be
found in Ref.~\cite{Kniehl:2005mk}.

We now describe our calculation of the differential cross section
$d^2\sigma/(dp_Tdy)$ of $p+\bar{p}\to X_c+X$, where
$X_c=D^0,D^+,D^{*+},D_s^+$ and $X$ comprises the residual final-state hadrons,
at NLO in the GM-VFN scheme.
A crucial ingredient entering this calculation are the non-perturbative FFs
for the transitions $a\to X_c$, where
$a=g,u,\overline{u},d,\overline{d},s,\overline{s},c,\overline{c}$.
For $X_c=D^{*+}$, such FFs were extracted at LO and NLO in the
$\overline{\rm MS}$ factorization scheme with $n_f=5$ massless quark flavors
several years ago \cite{BKK} from the scaled-energy ($x$) distributions
$d\sigma/dx$ of the cross section of $e^++e^-\to D^{*+}+X$ measured by the
ALEPH \cite{ALEPH} and OPAL \cite{OPAL1} Collaborations at CERN LEP1.
Two of us \cite{Kniehl:2005de} recently extended the analysis of
Ref.~\cite{BKK} to include $X_c=D^0,D^+,D_s^+,\Lambda_c^+$ by exploiting
appropriate OPAL data \cite{OPAL2}.
Besides the total $X_c$ yield, which receives contributions from
$Z\to c\bar{c}$ and $Z\to b\bar{b}$ decays as well as from light-quark and
gluon fragmentation, the ALEPH and OPAL Collaborations separately specified
the contribution due to tagged $Z\to b\bar{b}$ events yielding $X_b$ hadrons,
which then weakly decay to $X_c$ hadrons.
The contribution due to the fragmentation of primary $c$ quarks into $X_c$
hadrons approximately corresponds to the difference of these two measured 
distributions.
To test the scaling violations of these FFs and also the separation of the
$c\to X_c$ component, these FFs were employed to interpret the $x$
distributions of $e^++e^-\to X_c+X$ for center-of-mass energy
$\sqrt{S}=10.55$~GeV measured by the CLEO Collaboration \cite{Cleo} at LEPP
CESR, with very encouraging results.
Further details may be found in Ref.~\cite{Kniehl:2005de}.

In Refs.~\cite{BKK,Kniehl:2005de}, the starting scales $\mu_0$ for the DGLAP
evolution of the $a\to X_c$ FFs in the factorization scale $\mu_F^\prime$ are
taken to be $\mu_0=2m$, with $m=1.5$~GeV, for
$a=g,u,\overline{u},d,\overline{d},s,\overline{s},c,\overline{c}$ and
$\mu_0=2m_b$, with $m_b=5$~GeV, for $a=b,\overline{b}$.
The FFs for $a=g,u,\overline{u},d,\overline{d},s,\overline{s}$ are assumed to
be zero at $\mu_F^\prime=\mu_0$ and are generated through the DGLAP evolution
to larger values of $\mu_F^\prime$. 
Since the effect of the gluon FF is important at Tevatron energies, as was
found for $D^{*+}$ production in Ref.~\cite{Kniehl:2004fy}, we repeated the
fits of the $X_c$ FFs for the choice $\mu_0=m,m_b$.
This changes the $c$-quark FFs only marginally, but has a strong effect on
the gluon FF.
For shortage of space, these new FFs will be presented elsewhere.

The calculation of the cross section $d^2\sigma/(dp_Tdy)$ proceeds as outlined
in Ref.~\cite{Kniehl:2004fy}.
The full cross section consists of three contributions.
The first one contains all the channels with only gluons or light quarks in
the initial state and $c$-quark fragmentation.
Only this contribution carries explicit $m$ dependence.
Second, this contribution must be extended by allowing for $c$ quarks in the
initial state.
The third contribution is due to gluon or light-quark fragmentation.
The second and third contributions are calculated in the ZM-VFN scheme using
the hard-scattering cross sections derived in Ref.~\cite{ACGG}.
A certain part of these contributions is due to Feynman diagrams with internal
$c$-quark lines; another one is due to diagrams with external $c$-quark lines
and contains $m$-dependent logarithms, which are resummed.
Thus, in the FFN scheme, the $m$ dependence of these contributions would only
enter beyond NLO, which is reflected in the ZM-VFN scheme by the generic
suppression of the $c$-quark PDF relative to the gluon and light-quark ones
and of the gluon and light-quark FFs relative to the $c$-quark one.
This entitles us to omit this $m$ dependence by calculating the
$c$-quark-initiated contributions and those involving the fragmentation of
gluons or light quarks in the ZM-VFN scheme.
It turns out that the light-quark fragmentation contributions are negligible.
However, gluon fragmentation contributes approximately 40\% to the cross
section, almost independent of $p_T$.
For the $D^{*+}$ case, we showed in Ref.~\cite{Kniehl:2004fy} that the effect
of the $m$-dependent terms is much reduced in the full cross section, since
those parts that have to be calculated with $m=0$ dominate.
In fact, this observation carries over to the other charmed mesons considered
here.

We are now in a position to present our numerical results for the cross
sections of inclusive $D^0$, $D^+$, $D^{*+}$, and $D_s^+$ hadroproduction to
be directly compared with the CDF data \cite{CDF}, which come as distributions
$d\sigma/dp_T$ at $\sqrt{S}=1.96$~TeV with $y$ integrated over the range
$|y|\le1$.
For each $X_c$ species, the particle and antiparticle contributions are
averaged.
We work in the GM-VFN scheme with $n_f=4$, thus excluding $X_c$ hadrons from
$X_b$-hadron decays, which are vetoed in the CDF analysis \cite{CDF}.
We set $m=1.5$~GeV and evaluate $\alpha_s^{(n_f)}(\mu_R)$, where $\mu_R$ is
the renormalization scale, with $\Lambda^{(4)}_{\overline{\rm MS}}=328$~MeV
\cite{CTEQ6M}, corresponding to $\alpha^{(5)}(m_Z)=0.1181$.
We employ proton PDF set CTEQ6.1M from the CTEQ Collaboration \cite{CTEQ6M}
and the FFs introduced above.
We distinguish between the initial- and final-state factorization scales,
$\mu_F$ and $\mu_F^\prime$, so that we have three unphysical mass scales
altogether.
Our default choice is $\mu_R=\mu_F=\mu_F^\prime=m_T$, where
$m_T=\sqrt{p_T^2+m^2}$ the transverse mass.
In order to conservatively estimate the theoretical error due to the scale
uncertainty, we independently vary the values of $\mu_R/m_T$, $\mu_F/m_T$, and
$\mu_F^\prime/m_T$ between 1/2 and 2, and determine the maximum upward and
downward deviations from our default predictions.

\begin{figure*}[t]
\begin{center}
\begin{tabular}{llll}
{\parbox{3.5cm}{
\hspace*{-1.5cm}
\epsfig{file=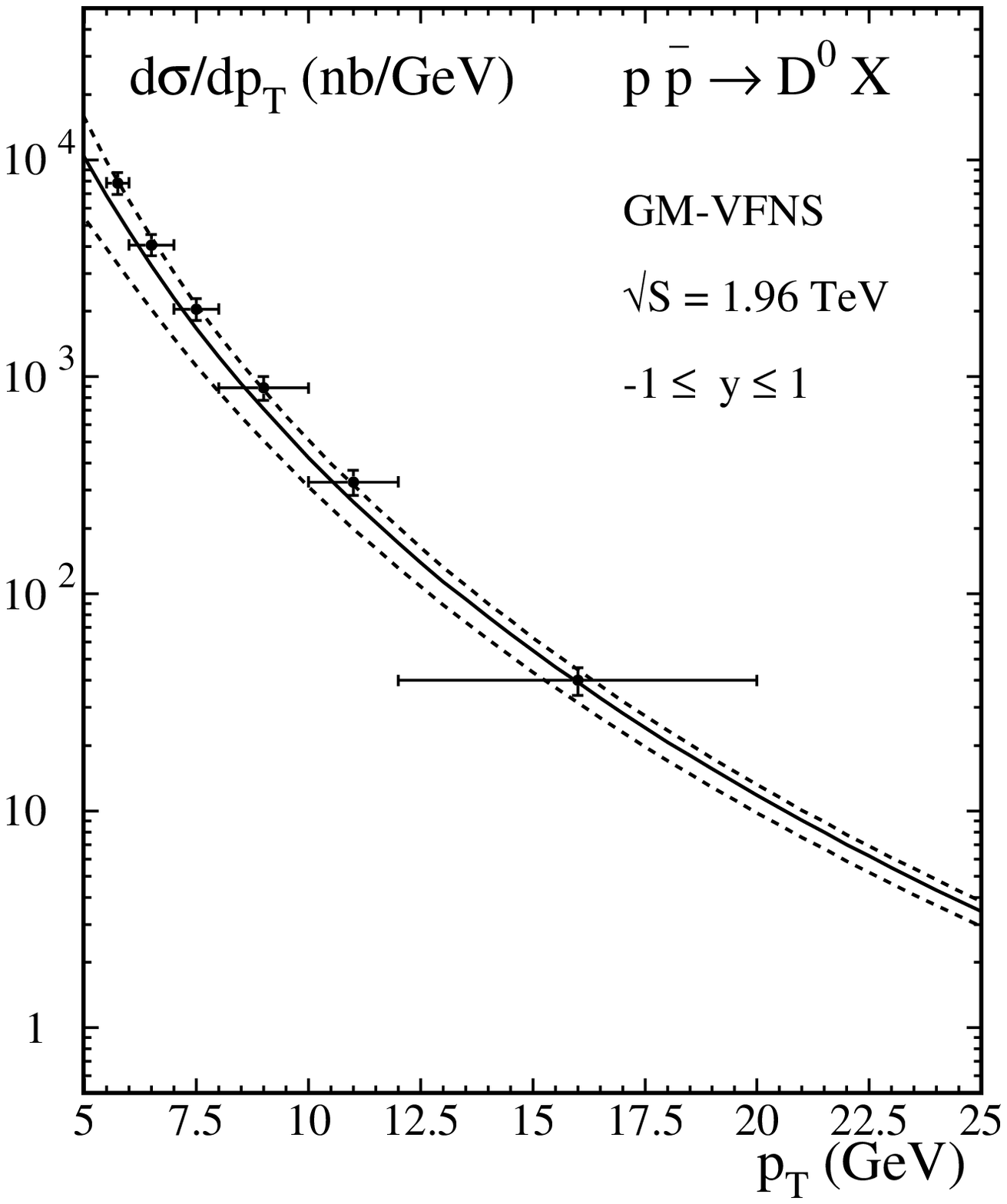,width=5.0cm}
}} &
{\parbox{3.5cm}{
\hspace*{-1.0cm}
\epsfig{file=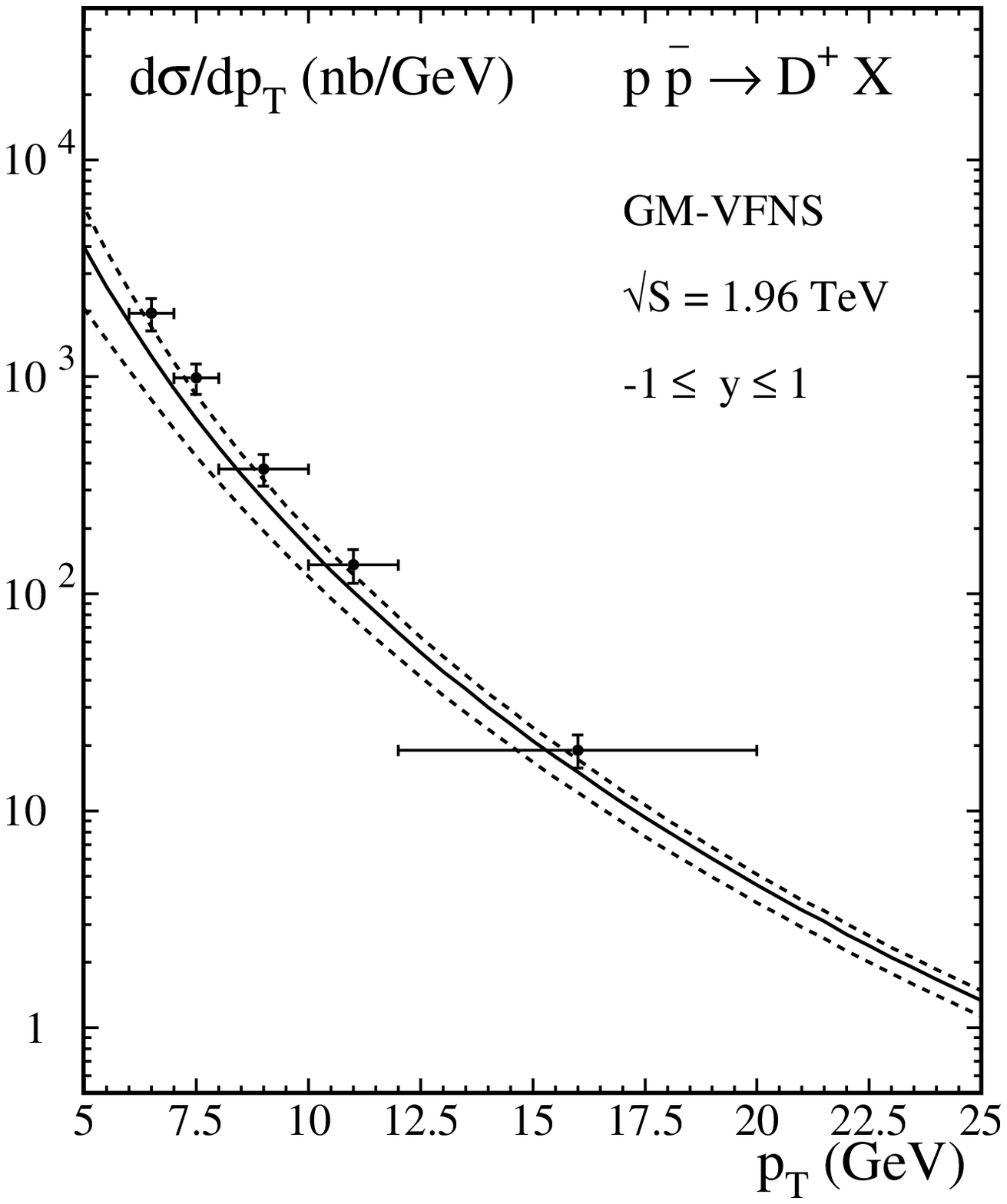,width=5.0cm}
}} &
{\parbox{3.5cm}{
\hspace*{-0.5cm}
\epsfig{file=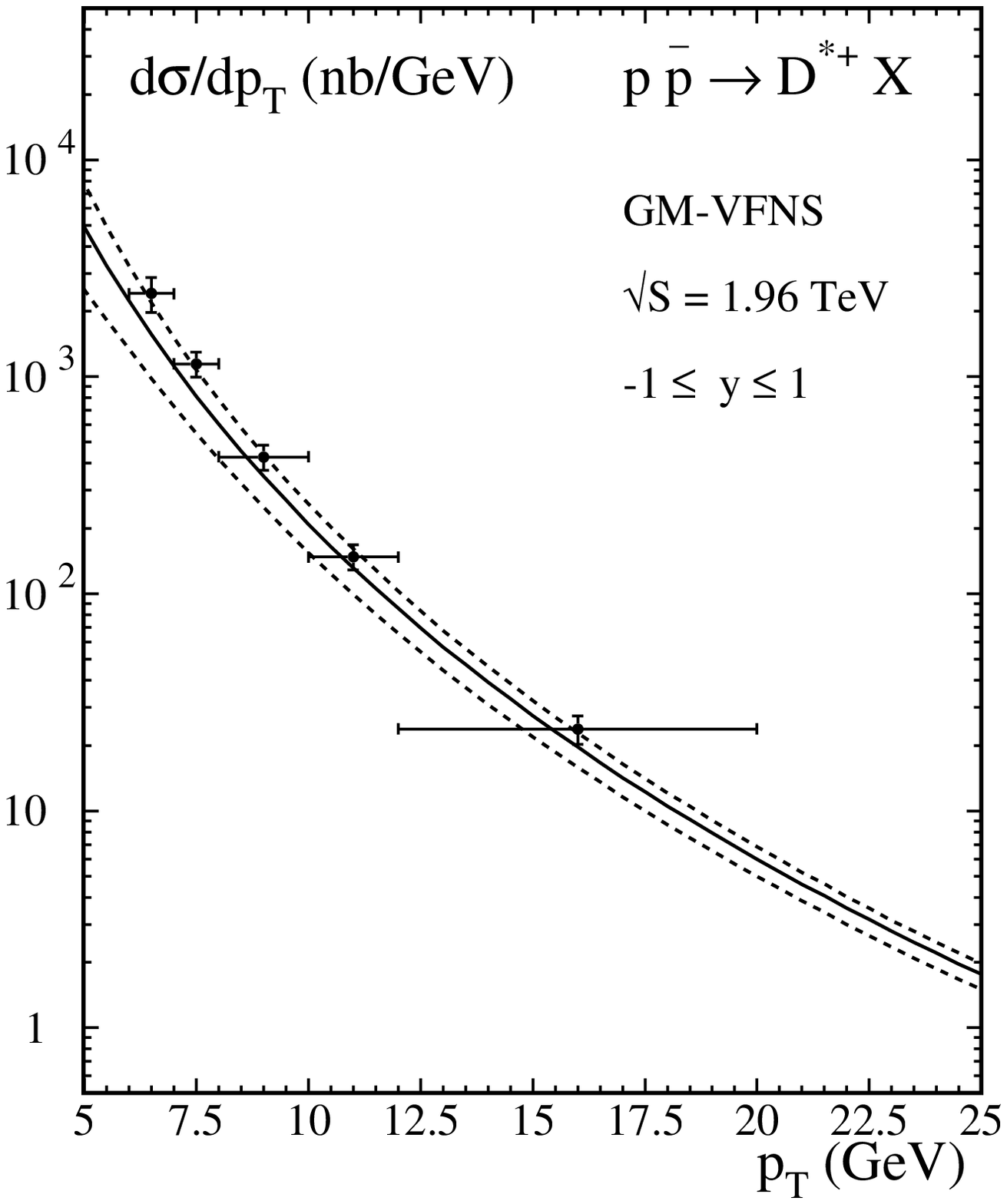,width=5.0cm}
}} &
{\parbox{3.5cm}{
\hspace*{-0.0cm}
\epsfig{file=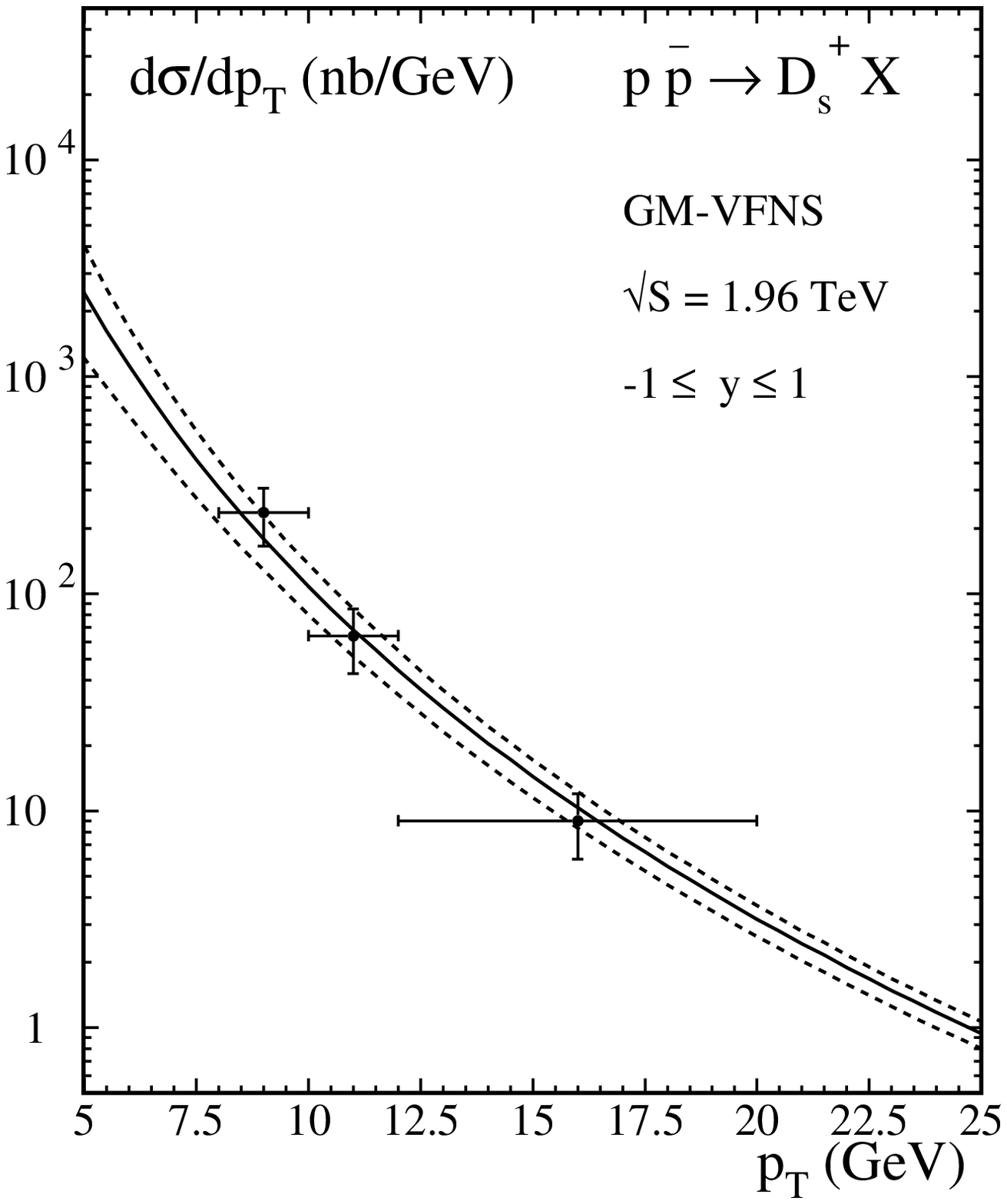,width=5.0cm}
}}
\end{tabular}
\end{center}
\vspace*{-0.75cm}
\caption{
Comparison of the CDF data \cite{CDF} with our NLO predictions for
$X_c=D^0,D^+,D^{*+},D_s^+$.
The solid lines represent our default predictions, while the dashed lines
indicate the scale uncertainty (see text).
}
\label{fig:fig1}
\end{figure*}
\begin{figure*}[t]
\begin{center}
\begin{tabular}{llll}
{\parbox{3.5cm}{
\hspace*{-1.5cm}
\epsfig{file=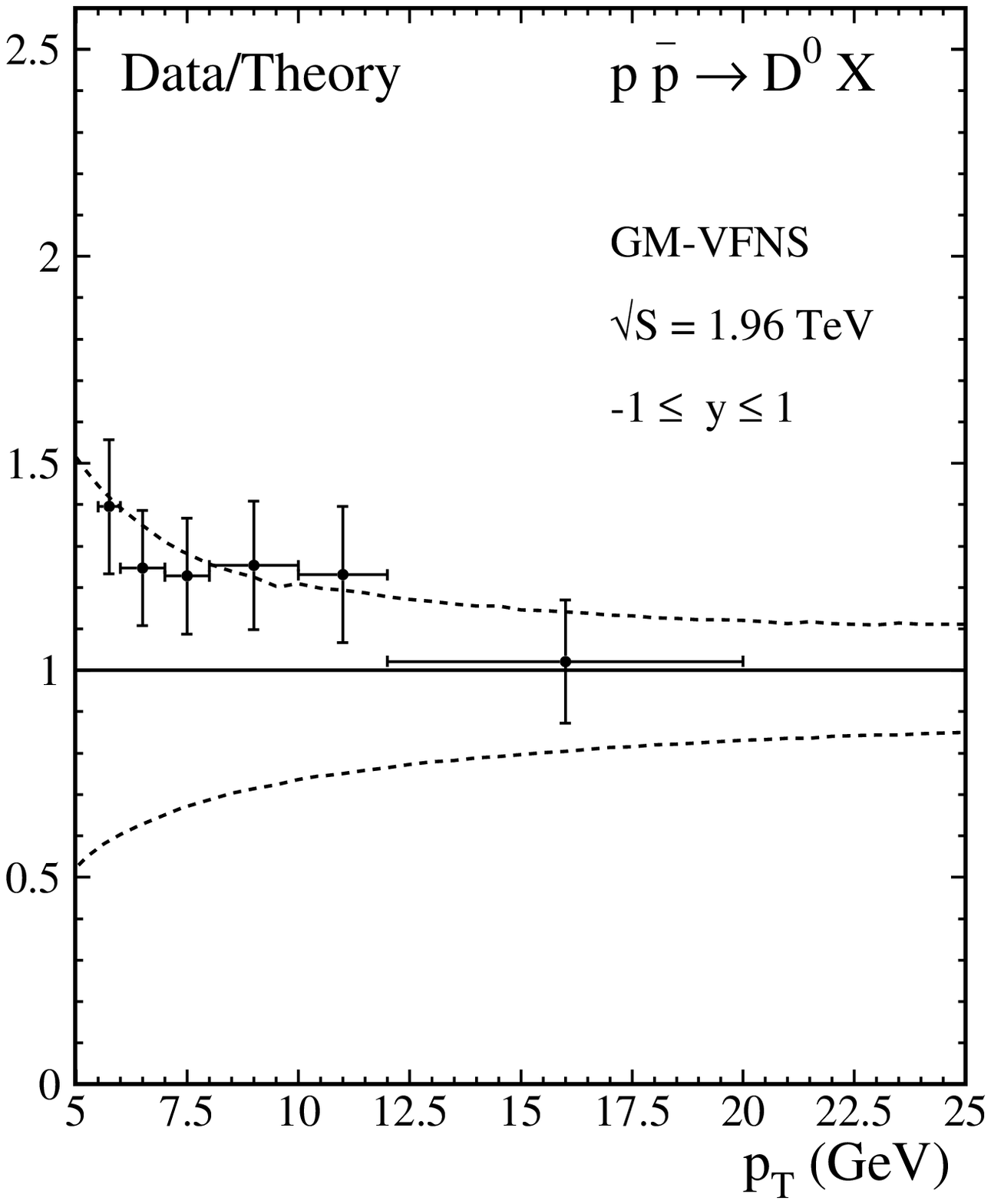,width=5.0cm}
}} &
{\parbox{3.5cm}{
\hspace*{-1.0cm}
\epsfig{file=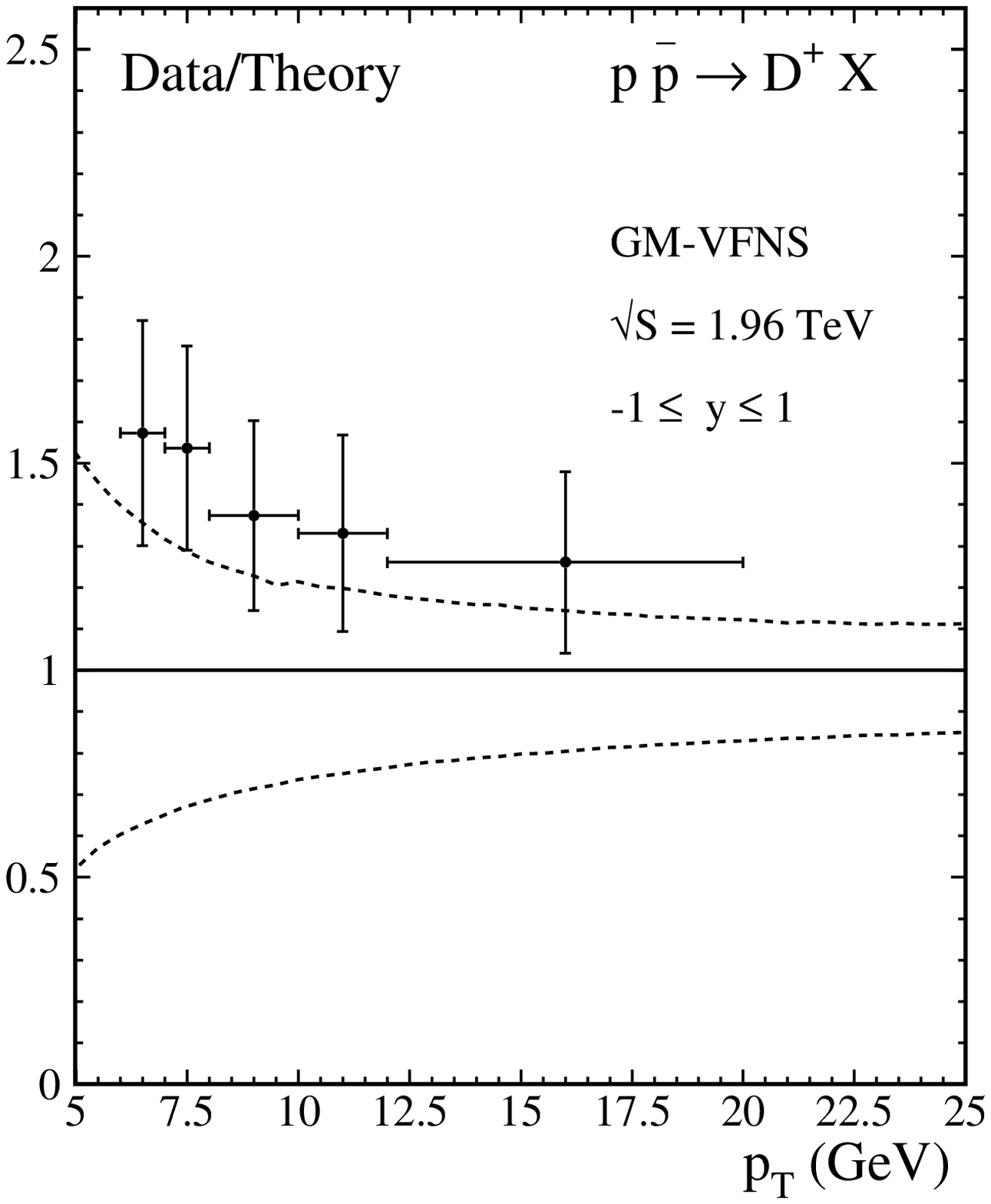,width=5.0cm}
}} &
{\parbox{3.5cm}{
\hspace*{-0.5cm}
\epsfig{file=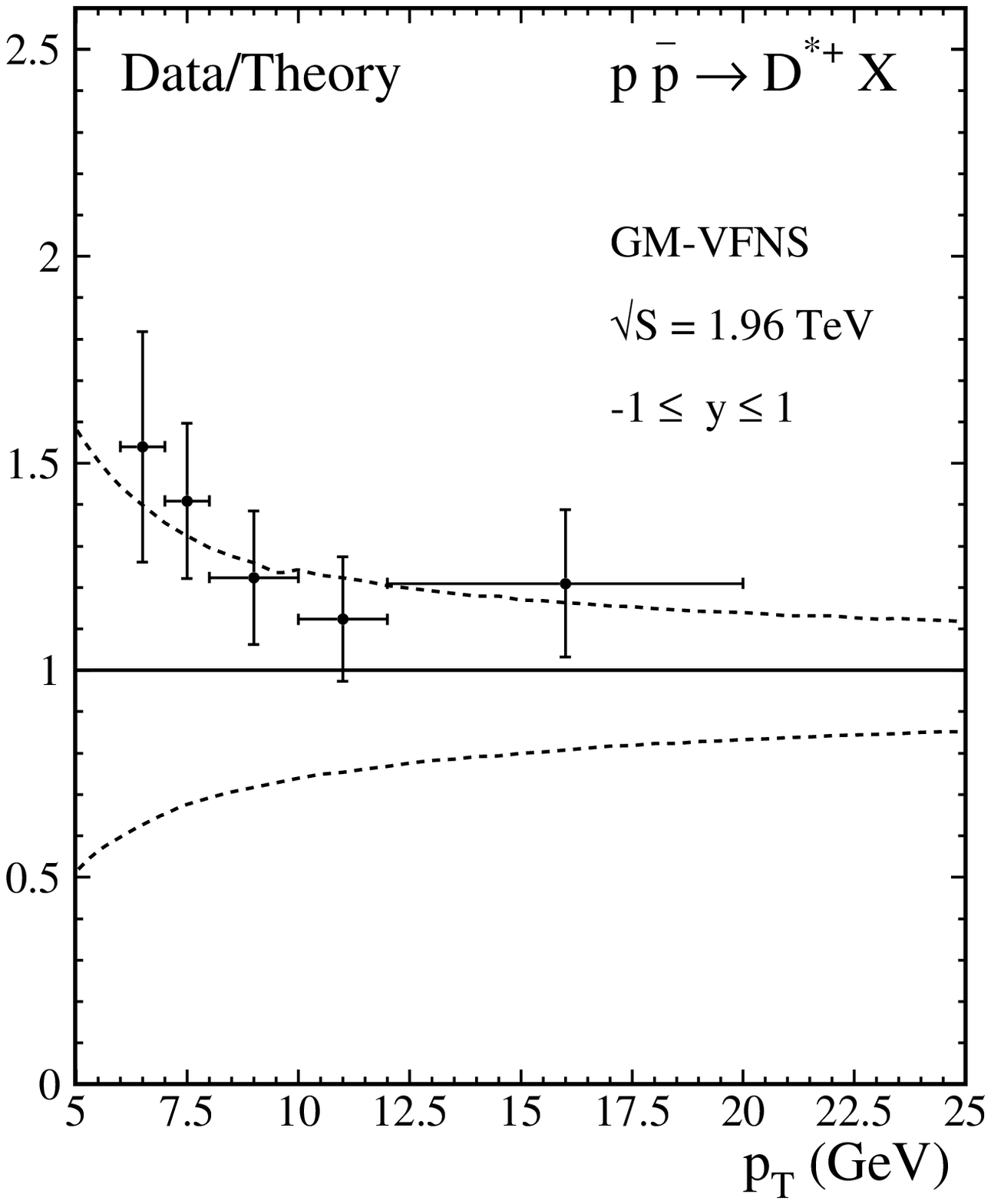,width=5.0cm}
}} &
{\parbox{3.5cm}{
\hspace*{-0.0cm}
\epsfig{file=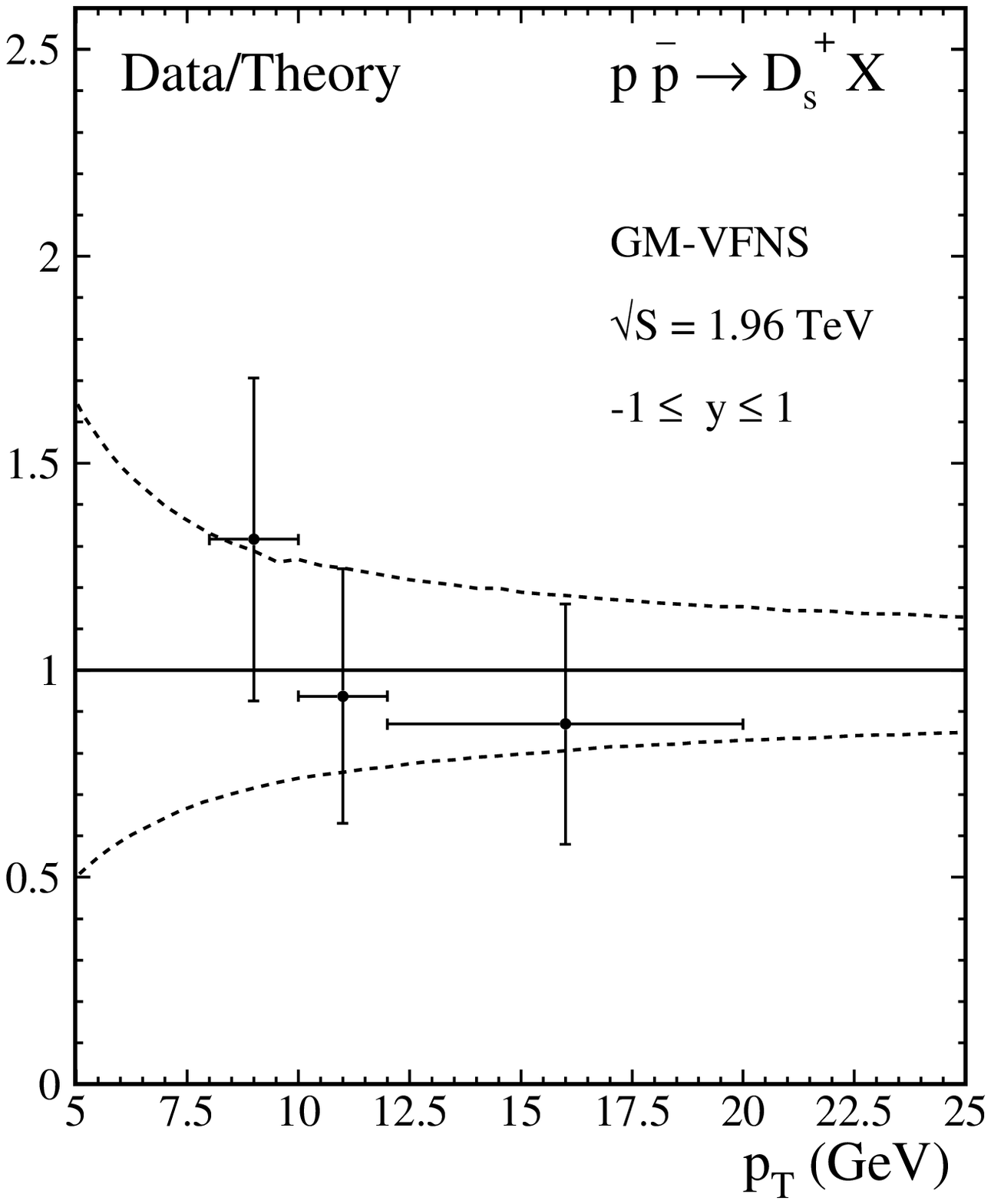,width=5.0cm}
}}
\end{tabular}
\end{center}
\vspace*{-0.75cm}
\caption{
Data-over-theory representation of Fig.~\protect\ref{fig:fig1} with respect to
our default predictions.
}
\label{fig:fig2}
\end{figure*}
Our theoretical predictions are compared with the CDF data on an absolute
scale in Fig.~\ref{fig:fig1} and in the data-over-theory representation
with respect to our default results in Fig.~\ref{fig:fig2}.
The four frames in each figure refer to $D^0$, $D^+$, $D^{*+}$, and $D_s^+$
mesons.
In all cases, we find good agreement in the sense that the theoretical and
experimental errors overlap, i.e.\ the notorious discrepancy between
experiment and theory \cite{CDF} mentioned in the introduction has disappeared.
In fact, our theoretical predictions provide the best description of the
CDF data obtained so far.

As for the $D^0$, $D^{*+}$, and $D_s^+$ mesons, many of the central data
points fall into the theoretical error band, while those for the $D^+$ mesons
lie somewhat above it.
With the exception of the $D_s^+$ case, the experimental results are gathered
on the upper side of the theoretical error band, corresponding to a small
value of $\mu_R$ and large values of $\mu_F$ and $\mu_F^\prime$, the $\mu_R$
dependence being dominant in the upper $p_T$ range.
As is evident from Fig.~\ref{fig:fig2}, in these cases, the central data
points tend to overshoot the central QCD predictions by a factor of about 1.5
at the lower end of the considered $p_T$ range, where the errors are largest,
however.
This factor is rapidly approaching unity as the value of $p_T$ is increased.
The tendency of measurements of inclusive hadroproduction in Tevatron run~II
to prefer smaller renormalization scales is familiar from single jets, which
actually favor $\mu_R=p_T/2$~\cite{DIS05}.

The overwhelming bulk of the theoretical error stems from the scale
uncertainty discussed above.
Residual sources of theoretical uncertainty include the variations in the
value of $m$ and the adopted PDF and FF sets.
We now quantitatively study the impact of these variations relative to the
typical example of our default prediction for $D^{*+}$ mesons.
The generous variation of $m$ by $\pm20\%$, from 1.2 to 1.8~GeV, induces a
shift in cross section of only $\pm 2\%$ at $p_T=5$~GeV, which rapidly
decreases towards larger values of $p_T$ because the $m$-dependent terms are
themselves reduced in size.
Switching to the NLO proton PDF set MRST2004 of Martin, Roberts, Stirling, and
Thorne \cite{MRST}, with $\Lambda^{(4)}_{\overline{\rm MS}}=347$~MeV and a
more physical parameterization of the gluon distribution to produce a better
description of the Tevatron inclusive jet data, produces a reduction ranging
from $-15\%$ at $p_T=5$~GeV to $-1\%$ at $p_T=20$~GeV.
The theoretical uncertainty due to the FFs was estimated in
Ref.~\cite{Kniehl:2004fy} to be of order 10\% or less in the $p_T$ range
considered here, by comparing FF sets \cite{BKK} fitted separately to slightly
incompatible ALEPH \cite{ALEPH} and OPAL \cite{OPAL1} data.

In conclusion, the GM-VFN scheme, which we elaborated at NLO for the inclusive
photo- \cite{KS} and hadroproduction \cite{Kniehl:2004fy,Kniehl:2005mk} of
heavy-flavored hadrons, resums large logarithms by the DGLAP evolution of
non-perturbative FFs and guarantees the universality of the latter as in the
ZM-VFN scheme and simultaneously retains the $m$-dependent terms of the FFN
scheme without additional theoretical assumptions.
Adopting this framework in combination with new fits of $D^0$, $D^+$,
$D^{*+}$, and $D_s^+$ FFs to OPAL data from LEP1 \cite{OPAL1,OPAL2}, we
managed for the first time to reconcile the CDF data on the production of
these mesons in Tevatron run~II \cite{CDF} with QCD within errors and thus
eliminated a worrisome discrepancy.
Furthermore, we presented the first NLO predictions for the $D_s^+$ data
\cite{CDF}.


\section*{Acknowledgments}
The work of I.S. was supported by DESY.
This work was supported in part by BMBF Grant No.\ 05 HT4GUA/4.




\begin{thebibliography}{1}
\bibitem{CDF} CDF Collaboration, D. Acosta {\it et al.},
Phys.\ Rev.\ Lett.\ {\bf91}, 241804 (2003).

\bibitem{BK} Calculation as in Ref.~\cite{KKP} with $D^{*+}$ FFs of
Ref.~\cite{BKK}.

\bibitem{KKP} B.A. Kniehl, G. Kramer, and B. P\"otter,
Nucl.\ Phys.\ {\bf B597}, 337 (2001).

\bibitem{BKK} J. Binnewies, B.A. Kniehl, and G. Kramer,
Phys.\ Rev.\ D {\bf58}, 014014 (1998).

\bibitem{CN} M. Cacciari and P. Nason,
JHEP {\bf0309}, 006 (2003).

\bibitem{ABE} CDF Collaboration, F. Abe {\it et al.},
Phys.\ Rev.\ D {\bf50}, 4252 (1994);
Phys.\ Rev.\ Lett.\ {\bf75}, 1451 (1995);
CDF Collaboration, D. Acosta {\it et al.},
Phys.\ Rev.\ D {\bf65}, 052005 (2002).

\bibitem{BER} E.L. Berger, B.W. Harris, D.E. Kaplan, Z. Sullivan,
T.M.P. Tait, and C.E.M. Wagner,
Phys.\ Rev.\ Lett.\ {\bf86}, 4231 (2001).

\bibitem{CG} M. Cacciari and M. Greco,
Nucl.\ Phys.\ {\bf B421}, 530  (1994).

\bibitem{BKNS} W. Beenakker, H. Kuijf, W.L. van Neerven, and J. Smith,
Phys.\ Rev.\ D {\bf40}, 54 (1989);
P. Nason, S. Dawson, and R.K. Ellis,
Nucl.\ Phys.\ {\bf B327}, 49 (1989); {\bf B335}, 260(E) (1990);
I. Bojak and M. Stratmann,
Phys.\ Rev.\ D {\bf67}, 034010 (2003).

\bibitem{CGN} M. Cacciari, M. Greco, and P. Nason,
JHEP {\bf9805}, 007 (1998).

\bibitem{KS} G. Kramer and H. Spiesberger,
Eur.\ Phys.\ J.\ C {\bf22}, 289 (2001);
{\bf28}, 495 (2003);
{\bf38}, 309 (2004).

\bibitem{Kniehl:2004fy} B.A. Kniehl, G. Kramer, I. Schienbein, and
H. Spiesberger,
Phys.\ Rev.\ D {\bf71}, 014018 (2005).

\bibitem{Kniehl:2005mk} B.A. Kniehl, G. Kramer, I. Schienbein, and
H. Spiesberger,
Eur.\ Phys.\ J.\ C {\bf41}, 199 (2005).

\bibitem{acot} M.A.G. Aivazis, J.C. Collins, F.I. Olness, and W.-K. Tung,
Phys.\ Rev.\ D {\bf50}, 3102 (1994).

\bibitem{Olness:1997yc} F.I. Olness, R.J. Scalise, and W.-K. Tung,
Phys.\ Rev.\ D {\bf59}, 014506 (1999).

\bibitem{ALEPH} ALEPH Collaboration, R. Barate {\it et al.},
Eur.\ Phys.\ J.\ C {\bf16}, 597 (2000).

\bibitem{OPAL1} OPAL Collaboration, K. Ackerstaff {\it et al.},
Eur.\ Phys.\ J. C {\bf1}, 439 (1998).

\bibitem{Kniehl:2005de} B.A. Kniehl and G. Kramer,
Phys.\ Rev.\ D {\bf71}, 094013 (2005).

\bibitem{OPAL2} OPAL Collaboration, G. Alexander {\it et al.},
Z.\ Phys.\ C {\bf72}, 1 (1996).

\bibitem{Cleo} CLEO Collaboration, R.A. Briere {\it et al.},
Phys.\ Rev.\ D {\bf62}, 072003 (2000);
CLEO Collaboration, M. Artuso {\it et al.},
{\it ibid.}\ {\bf70}, 112001 (2004).

\bibitem{ACGG} F. Aversa, P. Chiappetta, M. Greco, and J.Ph.\ Guillet,
Phys.\ Lett.\ B {\bf210}, 225 (1988); {\bf211}, 465 (1988);
Nucl.\ Phys.\ {\bf B327}, 105 (1989).

\bibitem{CTEQ6M} CTEQ Collaboration, J. Pumplin {\it et al.},
JHEP {\bf0207}, 012 (2002); 
CTEQ Collaboration, D. Stump {\it et al.},
{\it ibid.}\ {\bf0310}, 046 (2003).

\bibitem{DIS05} R. Field, for the CDF Collaboration,
in Proceedings of the XIII${}^{\rm th}$ International Workshop on Deep
Inelastic Scattering (DIS05), Madison, Wisconsin, 2005 (American Institute of
Physics, Melville, to be published);
B. Davies, for the D0 Collaboration, {\it ibid.}

\bibitem{MRST} A.D. Martin, R.G. Roberts, W.J. Stirling, and R.S. Thorne,
Phys.\ Lett.\ B {\bf604}, 61 (2004).

 
\end{thebibliography}
\end{document}